\documentclass[epj]{svjour}
\usepackage{amsmath}
\usepackage{amsmath}
\usepackage{amssymb}
\usepackage{amsfonts}
\usepackage{graphicx}
\usepackage{float}
\usepackage{wrapfig}
\usepackage{subfigure}
\usepackage{bm}
\usepackage{units}
\usepackage{grffile}
\usepackage{color}

\renewcommand{\Re}{\mathrm{Re}}
\newcommand{\Ro}{\mathrm{Ro}}
\newcommand{\Fr}{\mathrm{Fr}}

\begin{document}
\title{The spatio-temporal spectrum of turbulent flows}
\author{P. Clark di Leoni, P.J. Cobelli, and P.D. Mininni}
\institute{Departamento de F\'\i sica, Facultad de Ciencias
    Exactas y Naturales, Universidad de Buenos Aires and IFIBA, CONICET,
    Ciudad Universitaria, 1428 Buenos Aires, Argentina.}
\date{date/date}

\abstract{
    Identification and extraction of vortical structures and of waves
    in a disorganised flow is a mayor challenge in the study of
    turbulence. We present a study of the spatio-temporal behavior of
    turbulent flows in the presence of different restitutive
    forces. We show how to compute and analyse the spatio-temporal
    spectrum from data stemming from numerical simulations and from
    laboratory experiments. Four cases are considered: homogeneous and
    isotropic turbulence, rotating turbulence, stratified turbulence,
    and water wave turbulence. For homogeneous and isotropic
    turbulence, the spectrum allows identification of sweeping by the
    large scale flow. For rotating and for stratified turbulence, the
    spectrum allows identification of the waves, precise
    quantification of the energy in the waves and in the turbulent
    eddies, and identification of physical mechanisms such as Doppler
    shift and wave absorption in critical layers. Finally, in water wave
    turbulence the spectrum shows a transition from gravity-capillary
    waves to bound waves as the amplitude of the forcing is
    increased.
\PACS{{47.27.-i}{Turbulent flows} \and
    {47.35.-i}{Hydrodynamic waves} \and
    {68.03.Kn}{Dynamics (capillary waves)}}}
\maketitle

% Introduction 
\section{Introduction} 

Turbulence can be characterised as the disorganised
spa\-tio-tem\-po\-ral and chaotic evolution of a flow, by the
development of multi-scale structures, and in many cases by
intermittency (i.e., the development of localised energetic structures
in space and time, see e.g., \cite{Biferale04}). Although turbulence is
a spatio-temporal phenomena, limitations in experiments and in numerical
simulations led studies either towards spatial characterisation of the
flows (e.g., using two-point spatial structure functions, or the
wavenumber energy spectrum), or towards temporal characterisation (e.g.,
using the frequency spectrum of time series). On the one hand, computers
excel in the former approach, as velocity fields can be obtained today
in numerical simulations with relatively high spatial resolution, but
temporal cadence tends to be low as a result of the high computational
cost of writing large files in supercomputers. On the other hand,
experiments excel in the latter. Long time series are relatively easier
to obtain, while spatial resolution in the laboratory is often limited.

This led to different approaches to tackle the study of
turbulence. While some authors focused on spatial properties and
scaling laws, others considered the evolution of individual structures
such as vortex filaments or hairpin vortices
\cite{Hussain86,Rogers87}. Identification of individual (and coherent)
structures in the disorganised flow has always been a major quest for
turbulence research. Techniques such as the proper orthogonal
decomposition \cite{Berkooz93} allowed spa\-tio-tem\-po\-ral tracking
of these structures, and the generation of low dimensional models for
some turbulent flows \cite{Smith05}.

In the presence of restitutive forces (such as gravity and buoyancy in
a stratified fluid, or the Coriolis force in a rotating fluid), the flow
can also sustain waves that make the problem stiff: for strong enough
restitutive forces, waves introduce a fast time scale and their
amplitudes are slowly modulated by the evolution of the vortical
structures. The ability of waves to affect the properties of turbulent
flows has been recognised for a long time; as a few examples, it is
now clear they can alter the diffusion process in the ocean
\cite{Woods80}, make the flow anisotropic
\cite{Cambon89,Waleffe93,Cambon97}, and change the very nature of
nonlinear interactions \cite{Nazarenko}. Understanding the effect of
waves in turbulence has implications in geophysical, astrophysical,
and industrial flows. Although separating waves from eddies in a
turbulent flow has been deemed impossible in the past
\cite{Stewart69}, multiple advances allowed a formal treatment of
turbulence in the presence of waves
\cite{Cambon89,Waleffe93,Nazarenko}, as well as some ways to decompose
a turbulent flow into wave and vortical motions (see, e.g., 
\cite{Smith02,Bourouiba08,Sen12}). These decompositions are based on
spatial information and on the dispersion relation of the waves,
defining Fourier modes with zero wave frequency as vortical or
``slow'', and modes with non-zero wave frequency as waves or ``fast'',
with this approximation being valid only for very strong restitutive
forces.

The development of experimental techniques such as Particle Image
Velocimetry (PIV, see \cite{Adrian91}) or fringe projection
profilometry (including Fourier Transform Pro\-fi\-lo\-me\-try or FTP
\cite{Maurel09,Cobelli09}, and Empirical Mode Decomposition
Profilometry or EMDP \cite{Lagubeau15}), and the growth of computing
power (as well as the development of new technologies such as burst
buffers for the next generation of supercomputers \cite{Liu12}),
allowed obtaining experimental and numerical data with space and time
information. In this paper we will focus on recent methods developed
to detect and extract waves from the turbulent velocity field (and in
the case of water waves, also from the turbulent surface deformation
field), resulting from the superposition and nonlinear interaction of
waves and eddies.

Different methods can be employed to identify the presence of waves in a
turbulent flow, provided data in the spatial and temporal domain are
available. This is needed as waves are defined by their spatio-temporal
structure, i.e., by their dispersion relation. One approach, recently
used in experiments of rotating turbulence \cite{Campagne15}, is to
calculate the two-point spatial correlation of the temporal Fourier
transform of the velocity field, obtained from PIV measurements. In
\cite{Campagne15}, the authors were able to quantify anisotropy and to
identify inertial waves. Another approach is to use the fact that
bounded domains give rise to resonant frequencies and look for peaks in
the temporal Fourier spectrum, as was done in experiments of rotating
turbulence \cite{Bewley07,Rieutord12} in both cylindrical and square
containers (see also \cite{Lamriben11} for a recent study in a closed
container considering temporal information as well as the full spatial
fields). Yet another approach is to study the decorrelation time of
different spatial modes, and to see which are decorrelated by wave
dynamics (i.e., in one period of the waves) and which are decorrelated
by sweeping (i.e., in one turnover time of the large-scale flow). This
was done for magnetohydrodynamic (MHD) turbulence simulations
\cite{Servidio11} and for rotating turbulence simulations 
\cite{Favier10,Clark14a}. Using a similar technique, some authors looked
for peaks in the frequency spectrum of particular modes, as these should
be located at the frequencies corresponding to the wave dispersion
relation; these studies were conducted in MHD turbulence
\cite{Dmitruk09} and in stratified rotating turbulence
\cite{Lindborg07}. Finally, some observational studies of stratified
turbulence in the oceans attempted identification of waves by studying
Lagrangian trajectories in the fluid \cite{Dasaro00}.

Recently, computational and experimental advances made it possible to
calculate the complete spatio-temporal spectrum for all modes resolved
in an experiment or a simulation. This spectrum has been calculated,
e.g., in experiments \cite{Cobelli11,Aubourg15} and simulations
\cite{Clark14b} of water waves, in experiments of vibrating plates
\cite{Cobelli09,Yokoyama14}, in experiments \cite{Yarom14} and
simulations \cite{Clark14a} of rotating turbulence, in simulations of
magnetohydrodynamic turbulence \cite{Meyrand15}, in simulations of
stratified turbulence \cite{Clark15}, and in simulations of quantum
turbulence \cite{Nazarenko06a,Nazarenko06b,Clark15b} for flows of
superfluid helium or for Bose-Einstein condensates. These studies
allowed identification of wave modes, a precise quantification of how
much energy is present in these modes as a function of the
length scale,  and identification of physical effects associated with
the presence of waves.

Similar techniques were developed in other areas, specially in plasma
physics and in space physics. Experimental investigations of space
plasma turbulence have recently turned from single-point measurements
(which suffered invariably from ambiguities in disentangling temporal
and spatial variations) to multipoint measurements
\cite{Sahraoui10,Sahraoui11}.  A paradigmatic example of this is the
CLUSTER mission, which comprises an {\it in situ} investigation of the
Earth's magnetosphere using four identical spacecraft simultaneously
to distinguish between spatial and temporal variations
\cite{Escoubet}. The multi-spacecraft mission allows combination of
multiple time series recorded simultaneously at different points in space
to estimate the corresponding energy density in wavenumber-frequency
space \cite{Sahraoui11}. More recently, data from the Coronal 
Multi-channel Polarimeter (CoMP) allowed reconstruction of the
spatio-temporal spectrum and identification of Alfv\'en waves in the
solar corona \cite{Morton15}.

In the following sections we show how to compute and analyse the
spatio-temporal spectrum of turbulent flows, from numerical
simulations and experiments. We focus on four cases. Two were reported
extensively in \cite{Clark14a,Clark15} and are briefly summarised
here, and correspond respectively to numerical studies of rotating
turbulence, and of stratified turbulence. The spatio-temporal spectrum
allows identification of the waves, quantification of the energy in
the waves and in the turbulent eddies, and identification of physical
mechanisms such as Doppler shift and critical layer absorption. The
two other examples are new. We present the spatio-temporal spectrum of
numerical simulations of homogeneous and isotropic turbulence, and the
spatio-temporal spectrum of laboratory experiments of water wave
turbulence. In the former case, the analysis allows identification of
sweeping of the small scale vortices by the large scale flow. In the
latter, we show a transition from gravity-capillary wave turbulence to
bound waves as the amplitude of the forcing is increased.

% Methods 
\section{Methods}

% Numerical simulations
\subsection{Numerical simulations} 

To compute spatio-temporal resolved spectra in three-di\-men\-sions
(3D), we will consider data stemming from direct numerical simulations
of isotropic, rotating, and stratified turbulence. In the most general
case, the evolution of a rotating and stratified fluid in the
Boussinesq approximation is described by
\begin{align}
    \frac{\partial \mathbf{u}}{\partial t}
    +\mathbf{u}\cdot\mathbf{\nabla} \mathbf{u} - \nu \nabla^2 \mathbf{u}
    &=  - f \hat{z} \times \mathbf{u}- N \theta \hat{z} - {\bf \nabla} p +
    \mathbf{F},
    \label{NS}
    \\
    \frac{\partial \theta}{\partial t}
    +\mathbf{u}\cdot\mathbf{\nabla} \theta - \kappa \nabla^2 \theta &= 
    N \mathbf{u}\cdot \hat{z},
    \label{theta}
\end{align}
along with the incompressibility condition
\begin{align}
    \nabla \cdot \mathbf{u}=0 ,
\end{align}
where $\mathbf{u}$ is the velocity field, $\theta$ is the potential
temperature fluctuations, $p$ is the pressure (including
the centrifugal contribution), $f = 2\Omega$ (where $\Omega$ is the
rotation frequency, and the axis of rotation is in the $\hat{z}$
direction), $N$ is the Brunt-V\"ais\"al\"a frequency (note that
stratification is also in the $\hat{z}$ direction), $\mathbf{F}$ is an
external mechanical forcing, $\nu$ is the kinematic viscosity, and
$\kappa$ the thermal diffusivity (for simplicity we take $\kappa=\nu$,
i.e., a Prandtl number $Pr=\nu/\kappa = 1$). By linearising the
equations, it is straightforward to verify that inertia-gravity waves
are solutions to these equations, with dispersion relation
\begin{align}
    \omega (\mathbf{k}) = \pm \frac{\sqrt{
            N^2 k^2_\perp + f^2 k^2_\parallel}}{k} ,
    \label{reldisp}
\end{align}
where $k_\perp = (k^2_x+k^2_y)^{1/2}$, $k_\parallel = k_z$, and 
$k=(k^2_\perp+k^2_\parallel)^{1/2}$.

\begin{table}
    \begin{center}
        \begin{tabular}{| c | c | c | c |}
            \hline
            Simulation & Re & Fr & Ro
            \\
            \hline
            Isotropic turbulence & 5000 & -- & --
            \\
            Rotating turbulence & 5000 & -- & 0.015
            \\
            Stratified turbulence & 9700 & 0.01 & --
            \\
            \hline
        \end{tabular}
    \end{center}
    \caption{Dimensionless parameters for the three different
      numerical simulations. Re is the Reynolds number, Fr is the
      Froude number, and Ro is the Rossby number.}
    \label{adims}
\end{table}

All cases considered can be recovered from these equations. The
incompressible Navier-Stokes equation used for isotropic and
homogeneous turbulence is obtained from Eq.~(\ref{NS}) for 
$N=f=0$. In this case there are no waves, and only vortical 
structures are present in the flow. Equation (\ref{theta}) then
reduces to the equation for a passive scalar.

The purely rotating case is obtained from Eq.~(\ref{NS}) for $N=0$. As
the only restitutive force is the Coriolis force, the system can
sustain inertial waves, which from Eq.~\eqref{reldisp} follow the
dispersion relation given by
\begin{align}
    \omega_R (\mathbf{k}) = \pm \frac{f k_\parallel}{k} .
    \label{reldispR}
\end{align}

Finally, the purely stratified flow is obtained for $f=0$. Equation
\eqref{reldisp} now becomes the dispersion relation of internal gravity
waves,
\begin{align}
    \omega_S (\mathbf{k}) = \pm \frac{N k_\perp}{k} .
    \label{reldispS}
\end{align}

To solve these equations we use GHOST
\cite{Gomez05a,Gomez05b,Mininni11}, a parallel 3D pseudospectral code
with periodic boundary conditions, which uses either a second or fourth
order Runge-Kutta method for the time evolution. A spatial resolution of
$512^3$ points in a regular grid is used in all cases. In all
simulations the fluid starts from rest, and a constant in time forcing
is applied (constant forcing is used to prevent introduction of external
timescales to the system, that may be visible in the time spectrum). The
systems are then allowed to reach a turbulent steady state. Once this
stage is reached, we let the systems run for at least 10 large-scale
turnover times, in order to get enough statistics on the slower
timescales of the system. For isotropic and homogeneous turbulence, and
for stratified turbulence, we use an isotropic and randomly generated
three dimensional forcing acting at $k=1$. For rotating turbulence we
use Taylor-Green forcing with $\mathbf{F}= f_0 (\sin x \cos y \cos z
\hat{x} - \cos x \sin y \cos z \hat{y})$. See \cite{Clark14a,Clark15}
for more details of the simulations of rotating and of stratified
turbulence.

The dimensionless parameters that describe these systems are the
Reynolds number $\Re = U_{rms}L/\nu$, where $U_{rms}$ is the
r.m.s. velocity and $L$ the energy-containing length scale, the Rossby
number $\Ro = U_{rms}/(f L)$ which measures the relative strength of
rotation, and the Froude number $\Fr=U_{rms}/(NL)$ which measures the
relative strength of stratification. The values of these
dimensionless parameters for the three simulations considered below
are shown in Table \ref{adims}.

% Experimental setup
\subsection{Experimental setup} 

\begin{figure}
    \centering
    \includegraphics[trim={2.5cm .8cm 1.2cm 1.8cm},clip,width=8.5cm]{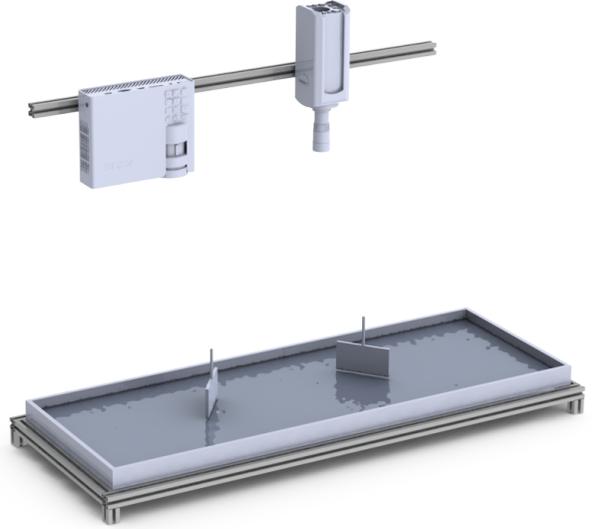}
    \caption{Experimental setup to generate water wave
      turbulence. The tank of $(200 \times 80)$ cm$^2$ is filled with
      water tainted with titanium dioxide to make it opaque and allow
      projection of a pattern on its surface. The two wavemakers
      driven by independent servomotors (not shown) excite random 
      perturbations of the free surface. A
      high resolution projector projects a fringe pattern of known
      characteristics (not shown in the scheme), and a fast speed
      camera captures the pattern deformation. The profilometry technique then 
      allows for reconstruction of the surface deformation with high
      temporal cadence and high spatial resolution
      \cite{Maurel09,Cobelli09,Lagubeau15}.}
    \label{diagram}
\end{figure}

The data to compute the spatio-temporal resolved spectrum of water
wave turbulence stems from an experimental setup to study waves in the
free surface of a liquid (see Fig.~\ref{diagram}). The setup consists of a
$(200\times80)$ cm$^2$ tank filled with water with depth at rest 
$h_0 = 5$ cm. Waves are generated in the tank by two piston-type wave
makers ($20$ cm large and $\approx 1$ cm immersed) independently driven
by two linear tubular servomotors. Two independent
random signals with  frequency range between 0 and $2.7$ Hz and with
amplitude $A$ are used to control the wavemakers. Three experiments
were performed, one with $A=1$ cm, another one with $A= 2$ cm, and a
third one with $A=3$ cm. Under this set of conditions, and in the
absence of wave breaking, waves should follow the linear dispersion
relation of gravity-capillary waves
\begin{align}
    \omega_W (k) =  \sqrt{\tanh(h_0 k) \left( g k + \frac{\gamma}{\rho}
            k^3 \right)},
    \label{reldispSW}
\end{align}
where $\gamma$ is the surface tension, and $\rho$ the density of water.

The surface height deformation $h(x,y,t)$ is obtained with high temporal
cadence and high spatial resolution using a fringe projection
profilometry technique \cite{Maurel09,Cobelli09}. Titanium dioxide is
added to the water as dye, in order to render the free surface 
light diffusive without changing significatively its rheological properties
\cite{Przadka2012} and be able to project onto it a controlled pattern by
means of a high-resolution, high-contrast projector. A
fast speed camera, with a resolution of $1024\times1024$ px$^2$ and
inspecting an area of $(41.6\times41.6)$ cm$^2$, is then used to capture
deformations of the pattern as the result of the surface deformation.
The size of the projected pixel, about $0.04$ cm, sets the spatial
resolution. The temporal resolution is $1/F$, where $F$ is the
acquisition frequency (in our case $F = 250$ Hz). The profilometry technique
reconstructs the height of the surface $h(x,y,t)$ from the distortion in
the patterns captured by the camera.

\begin{figure}
    \centering
    \includegraphics[width=8.5cm]{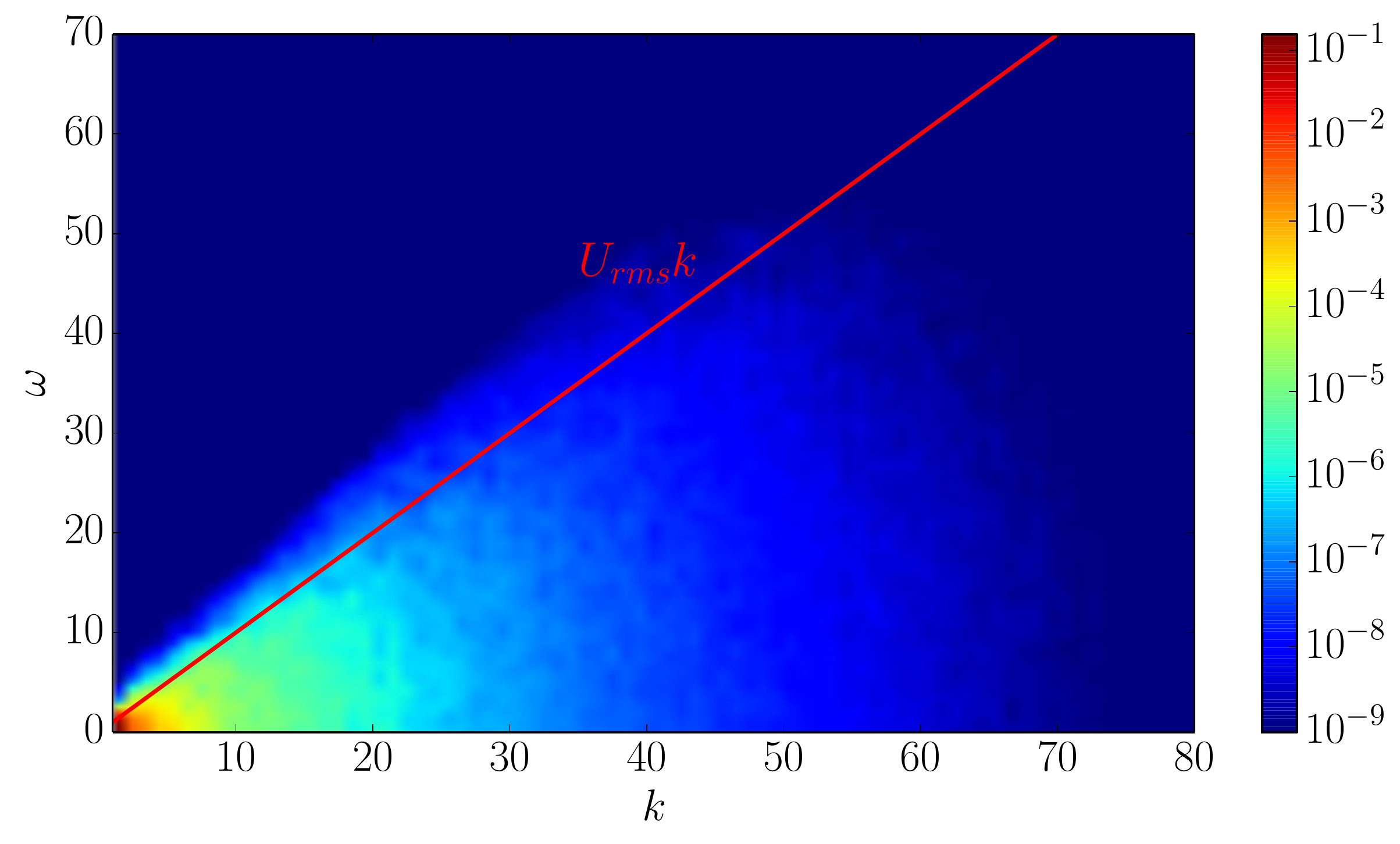}
    \caption{Spatio-temporal spectrum $E(k,\omega)$ in a numerical
      simulation of isotropic and homogeneous turbulence. The solid
      curve corresponds to $\omega =U_{rms} k$. As a result of 
      sweeping of the small scale eddies by the large scale flow,
      most of the energy is concentrated in the region $\omega \le
      U_{rms} k$.}
    \label{hitekw}
\end{figure}

\begin{figure}
    \centering
    \includegraphics[width=8cm]{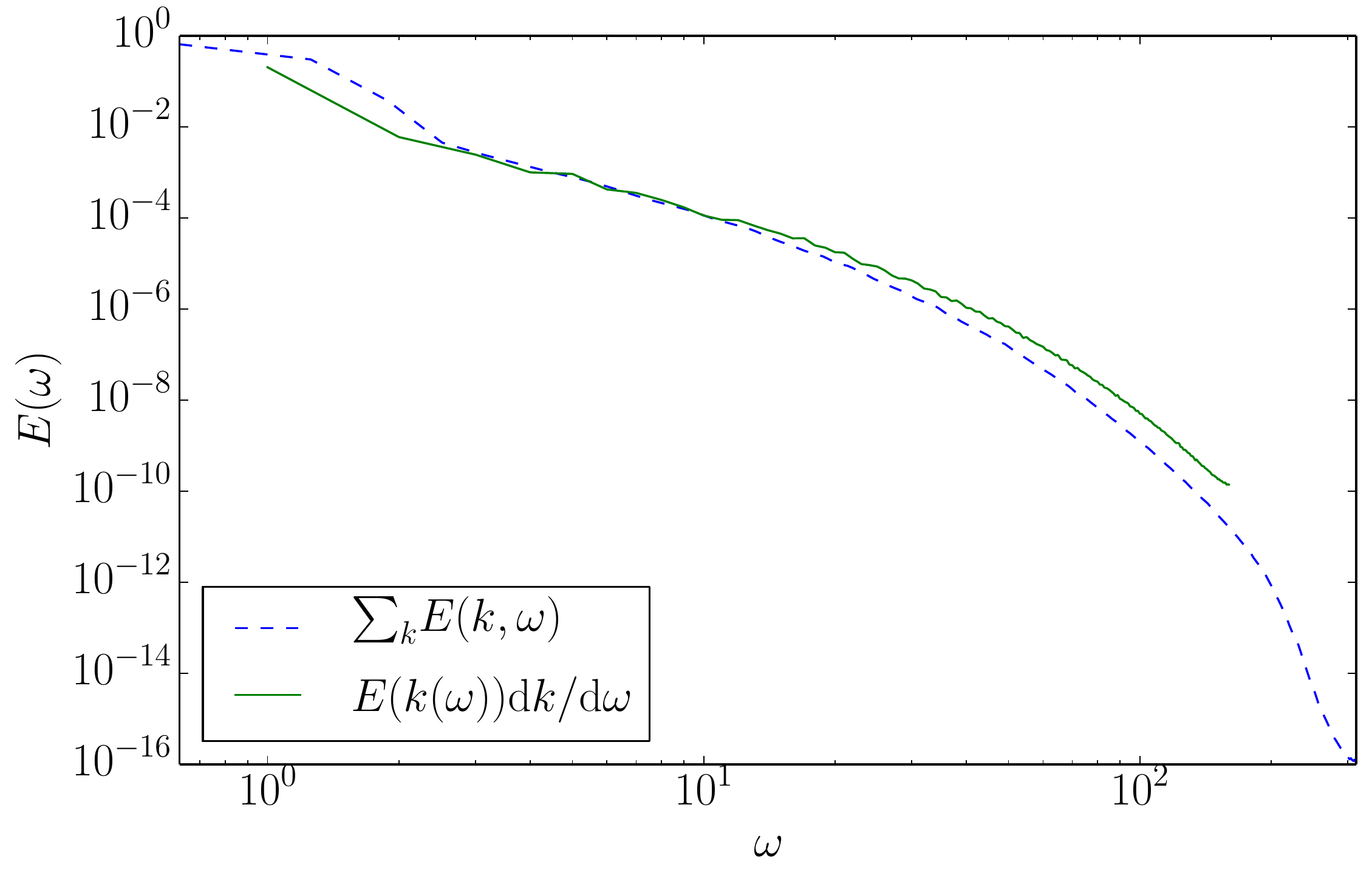}
    \caption{Frequency spectrum $E(\omega)$ for the homogeneous and
        isotropic turbulence simulation, calculated directly as
        $E(\omega) = \sum_k E(k,\omega)$ (dashed blue line). The solid
        (green) line is obtained by transforming the spatial spectrum
        $E(k)$ into a frequency spectrum $E(k(\omega))d\omega/dk$ using
        the sweeping relation $\omega=U_{rms}k$. As sweeping is the
        dominant Eulerian time scale in the system, one spectrum can
        be recovered from the other with reasonable agreement of both
        spectra in the inertial range.}
    \label{hitew}
\end{figure}

% Constructing the spatiotemporal spectra
\subsection{Construction of the spatio-temporal spectrum
 \label{sec:spectrum}}

In principle, computation of the spatio-temporal spectrum reduces to
computing Fourier transforms of the space and time resolved numerical
or experimental data. In practice, some provisions may be made to
allow for efficient storage and correct handling of the data.

First of all, both in numerical simulations and in experiments, the
acquisition frequency must be at least two times larger than the
frequency of the fastest waves one wants to study, and the total time
of acquisition should be larger than the period of the slowest waves
in the system, and larger than the turnover time of the slowest eddies. 

In numerical simulations storage constraints (both in space as well as
in I/O speed) make it difficult to store the velocity fields at all points
in space with high temporal cadence (e.g., every time step, or every a
very few time steps), specially at high spatial resolutions. As a
result, we store the Fourier transform of the velocity field 
$\hat{\bf u}({\bf k},t)$ with high temporal cadence, and for selected
Fourier modes ${\bf k}$. In most cases it suffices to store all Fourier
modes in three planes (corresponding to the planes with
$k_x=0$, $k_y=0$, and $k_z=0$). This allows reconstruction of the
spatio-temporal energy spectrum in three planes $E(k_x=0,k_y,k_z,\omega)$,
$E(k_x,k_y=0,k_z,\omega)$, and $E(k_x,k_y=0,k_z,\omega)$, where, e.g.,
the first is computed from the spatial Fourier modes of the velocity
field in the plane with $k_x=0$ as
\begin{align}
E(k_x=0,k_y,k_z,\omega) = \frac {1}{2} \left| \int{ 
    \hat{\bf u}(k_x=0,k_y,k_z,t) e^{-i \omega t} dt} \right|^2 .
\end{align}
Note that in all cases the computation results in a four
  dimensional spectrum, or in multiple three dimensional spectra that
  are better studied by plotting slices for constant values of the
  wavenumber Cartesian components or of the frequency. 
Furthermore, if the spectrum is isotropic (in the absence of external
forces) or axisymmetric (in the rotating or stratified cases), symmetry
considerations allow for reconstruction of the isotropic spectrum
$E(k,\omega)$ or of the axisymmetric spectrum
$E(k_\perp,k_\parallel,\omega)$ from these three spectra.
(see, e.g., \cite{Davidson}).

When the data is not periodic (as is the case with the spatial
  data in experiments, or with temporal data in both experiments and
  in simulations) it is advisable to use a window function to avoid
  introducing artifacts when the Fourier transforms are performed, and
  to mitigate spectral leakage. In the following, a flat top filter
  will be used when computing the temporal Fourier transforms for all
  the data coming from simulations. For the experimental data, a
  Hanning  window will be used in both space and time. Furthermore, 
  in experiments it is relatively easier to obtain long time signals;
  this allows us to perform a Welch average in order to reduce noise
  in the frequency spectra.

% Results 
\section{Results} 

\begin{figure}
    \centering
    \includegraphics[width=8.5cm]{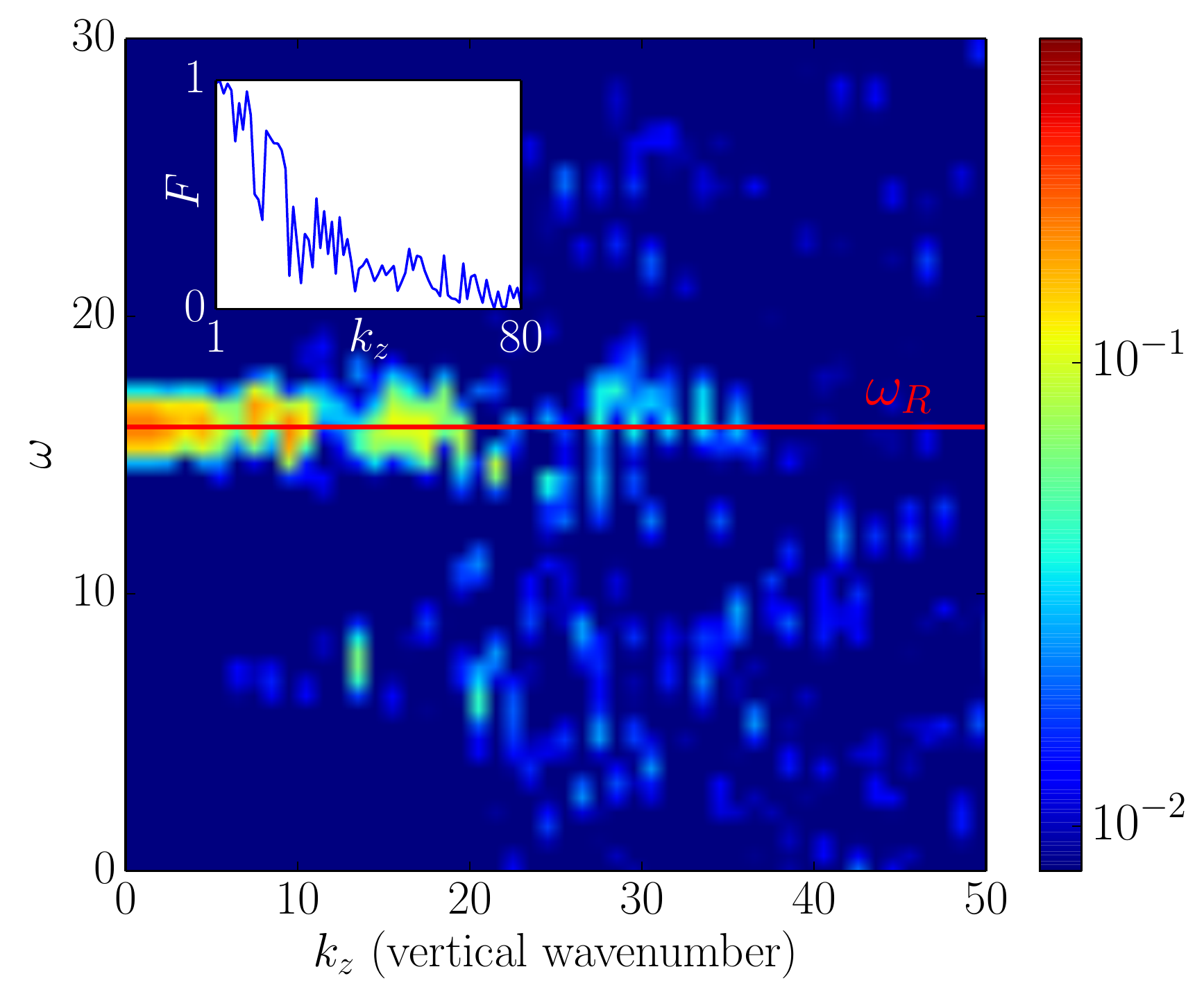}
    \includegraphics[width=8.5cm]{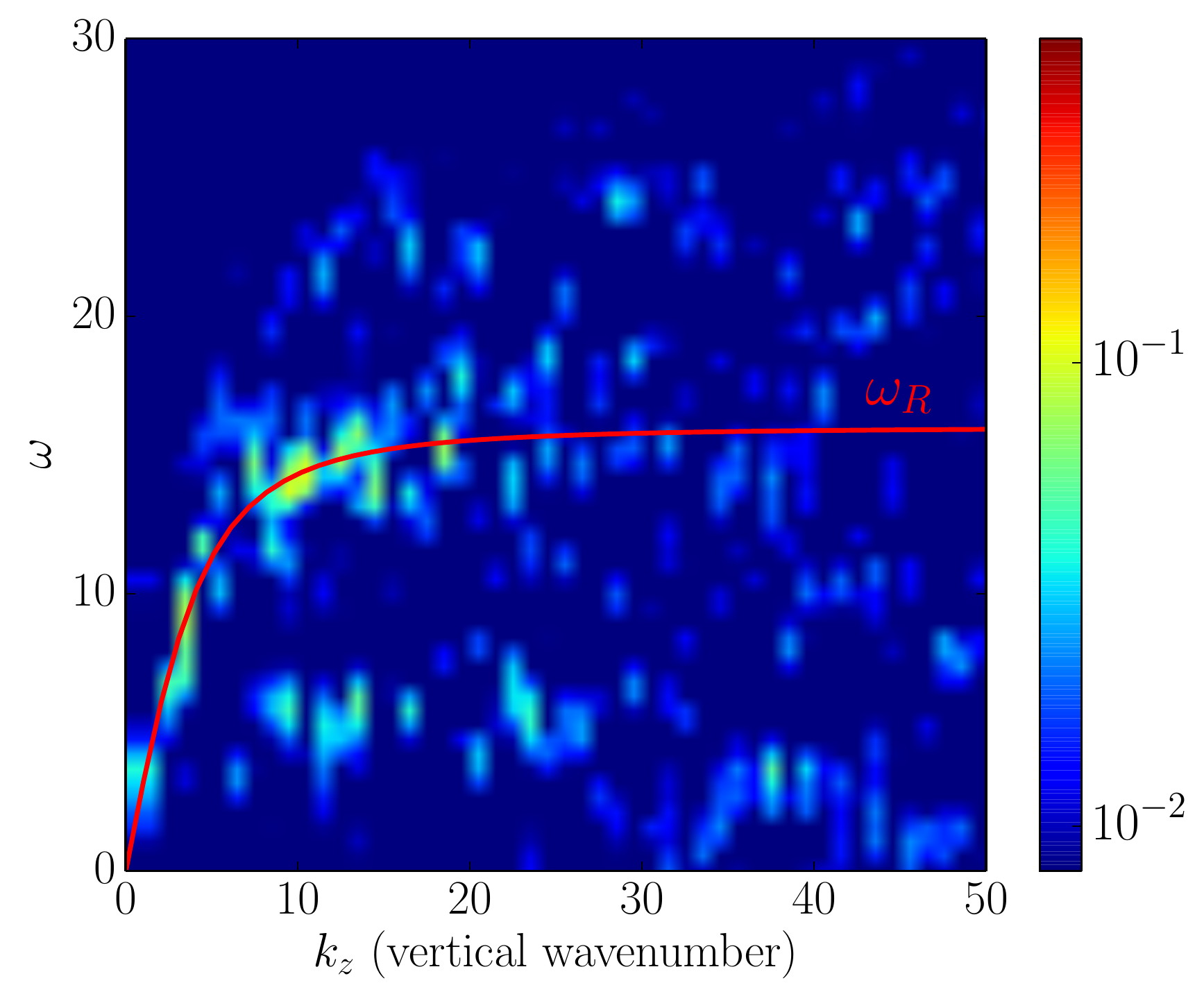}
    \caption{Spatio-temporal energy spectrum $E(k_x=0,k_y,k_z,\omega)$
      from numerical simulations of rotating turbulence. Two slices are
      shown for fixed values of $k_y$. {\it Top}: $k_y=0$, and {\it
      bottom}: $k_y=1$. The dispersion relation of inertial waves,
      $\omega_R({\bf k})$, is shown by the solid line. For
      small wavenumbers most of the energy is concentrated around this
      dispersion relation. {\it Inset}: Ratio of the energy in the
      vicinity of the modes satisfying the dispersion relation, to the
      total energy in the same wavenumber.}
    \label{rotekw}
\end{figure}

% Homogeneous and isotropic turbulence
\subsection{Homogeneous and isotropic turbulence\label{sec:HIT}}

In the absence of restitutive forces, an incompressible fluid cannot
sustain waves. Isotropic and homogeneous turbulence can then be
characterised by two timescales which follow from the Navier-Stokes
equation: the eddy turnover time at the scale $\ell$, given by
$\tau_\ell = \ell/u_\ell$ where $u_\ell$ is the characteristic
velocity of an eddy of size $\ell$, and the sweeping time, given by $T
= \ell/U_{rms}$, which describes the sweeping of vortices of
size $\ell$ by the flow at the largest scale. The former is related 
to the local interaction of modes in Fourier space, i.e., of eddies of
similar sizes, while the latter deals with the non-local interaction
in Fourier space resulting from the advection of small eddies by
the energy containing ones. It has been theorised
\cite{Tennekes75,Chen89} and later shown \cite{Nelkin90,Sanada92} that
the temporal decorrelation of Fourier modes of the Eulerian velocity
field in homogeneous and isotropic turbulence is determined by this
sweeping. In numerical simulations sweeping is often
quantified by computing the decorrelation time of individual Fourier
modes, using two-time correlation functions. The spatio-temporal
spectrum of isotropic and homogeneous turbulence has not been
considered so far to quantify its effect.

\begin{figure}
    \centering
    \includegraphics[width=8cm]{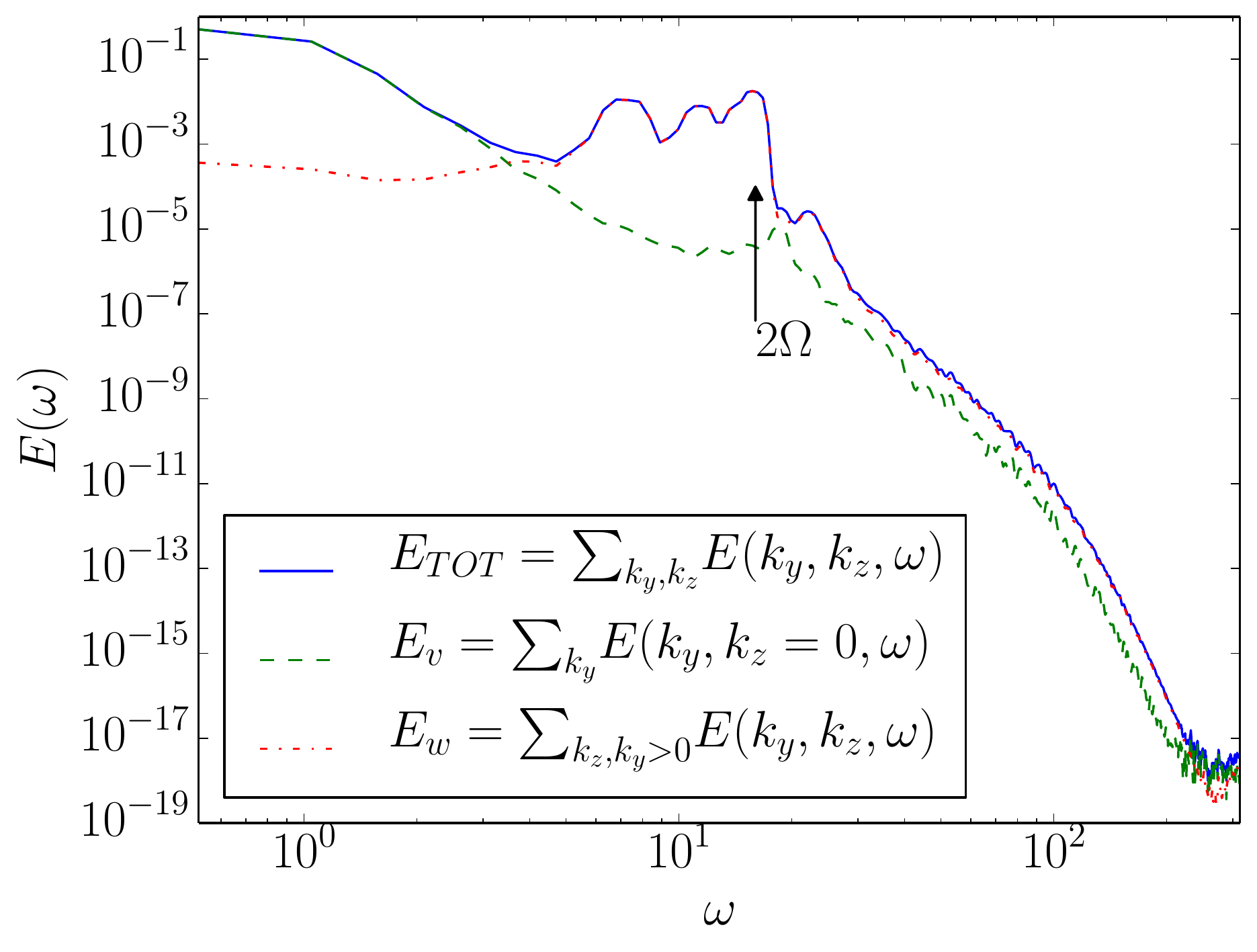}
    \caption{Frequency spectra of kinetic energy in the simulation of
      rotating turbulence. Here $E_v(\omega)$ (dashed green line) 
        has only the contribution from the vortical modes, while
        $E_w(\omega)$ (dash-dotted red line) has the contribution 
        of the so-called ``wave'' modes (i.e., modes with $k_z>0$).
        The total frequency spectrum $E_{TOT}$ 
        (solid blue line) is obtained by summing the spatio-temporal
        spectrum over all wavenumbers. The largest peak in $E_w$ 
        corresponds to a frequency $2\Omega$, indicated by the
        arrow. The contribution of the eddies dominate the smaller
        frequencies with $\omega\lesssim4$, where waves are too slow
        and the two-dimensional vortices are strong. For frequencies
        larger than $2\Omega$ modes cannot be described as inertial
        waves and eddies must dominate all spectra.}
    \label{rotew}
\end{figure}

Figure \ref{hitekw} shows the spatio-temporal energy spectrum computed
using the velocity field stemming from the simulation of isotropic and
homogeneous turbulence (see Table \ref{adims}). The thick red line
indicates the relation $\omega = U_{rms} k$, corresponding to the
frequency of sweeping by a flow with constant velocity
$U_{rms}$. Sweeping appears in the spectrum as the
concentration of energy along and below this line. Structures of size
$\ell \sim 1/k$ are advected by the large-scale velocity which
roughly fluctuates between 0 and $\approx U_{rms}$.  In other words, at
a given wavenumber $k$, all frequencies are excited up to  $\omega =
U_{rms} k$, and not just the frequency corresponding to the eddy
turnover time at that wavenumber, $\omega_k \sim 1/\tau_\ell$. Note
that as was pointed out in \cite{Tennekes75,Chen89}, the dominance of
the sweeping time as the Eulerian decorrelation time, over the
eddy turnover time, is needed for the frequency spectrum of isotropic
and homogeneous turbulence to be $E(\omega) \sim \omega^{-5/3}$,
provided the wavenumber spectrum is Komogorov's $E(k) \sim
k^{-5/3}$.

Indeed, the spatio-temporal spectrum in Fig.~\ref{hitekw} allows
    direct computation of both spectra, $E(\omega)$ and $E(k)$, without
    relying on extra assumptions such as the Taylor hypothesis. We can
    actually put this fact to good use to further verify that the
    dominant timescale in the system is the one given by the
    sweeping mechanism. In Fig.~\ref{hitew} we show $E(\omega)$
    computed as $E(\omega)=\sum_k E(k,\omega)$ (i.e., computing it
    directly from the spatio-temporal spectrum), compared with the
    spectrum $E(\omega(k)) d\omega/dk$ obtained from the spatial
    spectrum $E(k)$ using Taylor (or sweeping) hypothesis (i.e., from
    a change of variables using the relation $\omega=U_{rms}k$). The
    good agreement between the inertial ranges of the two spectra
    confirms that sweeping is the dominant time scale for the Eulerian
    velocity. Note also the presence of the well known bottleneck in
    the dissipative range of the spatial spectrum.

% Rotating turbulence
\subsection{Rotating turbulence} 

Rotation breaks down the isotropy of the flow, as the axis of rotation
establishes a preferred direction. The Coriolis term also acts as a
restitutive force that allows the fluid to sustain inertial
waves. For strong enough rotation, these waves are much faster than
the eddies, at least for a subset of all Fourier modes corresponding
to modes with $k_\parallel \neq 0$ (as for $k_\parallel = 0$ the
dispersion relation of the waves vanishes), and to modes with
$k<k_\Omega$ (where $k_\Omega$ is the Zeman wavenumber for which the
eddies become as fast as the waves and the flow recovers isotropy
\cite{Mininni12}). Non-linear resonant interactions between triads of
these waves then become the preferred energy transfer mechanism
\cite{Newell69,Cambon89,Waleffe93}. Given three modes with wave
vectors ${\bf k}$, ${\bf p}$, and ${\bf q}$, they can interact and
transfer energy if
\begin{align}
{\bf k} + {\bf p} + {\bf q} = 0 , \\
\omega_R({\bf k}) + \omega_R({\bf p}) + \omega_R({\bf q}) = 0 .
\end{align}
The last relation is the resonant condition. These relations dictate
that energy in a rotating flow is transferred preferentially towards
modes with smaller $k_\parallel$ \cite{Waleffe93} (as modes with
$k_\parallel=0$ trivially satisfy the resonant condition), resulting
in a growth of the anisotropy and the 
quasi-bidi\-men\-sion\-al\-iza\-tion of the flow.

Multiple wave turbulence theories have been proposed for rotating
turbulence based on these conditions (see, e.g.,
\cite{Cambon89,Cambon97,Cambon04}). However, the theories 
often assume the rapidly rotating limit in which waves overshadow the
eddies completely. Thus, the slow or vortical modes with 
$k_\parallel = 0$ and the modes with $k\approx k_\Omega$ cannot be
properly described within the framework of these theories. Moreover,
in simulations and experiments the discrimination between waves and
eddies often reduces to considering modes with 
$k_\parallel \approx 0$ (for which $\omega_R({\bf k})\approx0$) as
eddies, and all other modes as waves. Computation of the 
spa\-tio-tem\-po\-ral spectrum, which has been recently performed for
numerical simulations \cite{Clark14a} and for experimental data
\cite{Yarom14}, allows for a proper identification of the waves and a
quantification of their relevance at different scales.

The spatio-temporal spectrum of the kinetic energy
$E(k_x,k_y,k_z,\omega)$ for the simulation of rotating turbulence (see
Table \ref{adims}) is shown in Fig.~\ref{rotekw}, for $k_x=0$ and for
two fixed values of $k_y$ ($k_y=0$ and $k_y=1$). As mentioned in
Sec.~\ref{sec:spectrum}, compared with results previously shown in
\cite{Clark14a}, the spectrum in Fig.~\ref{rotekw} was computed using
a flat top window to reduce spectral leakage. Note that unlike the
spectrum in Fig.~\ref{hitekw}, sweeping effects cannot be observed in
these spectra. Instead, energy is accumulated near modes satisfying
the dispersion relation of inertial waves. In Fig.~\ref{rotekw} we
also show the ratio of the energy in these modes to the total energy
in the same wavenumber
\begin{equation}
F(k_z) = \frac{E(k_x=0,k_y=0,k_z,\omega=\omega_R)}
    {\int E(k_x=0,k_y=0,k_z,\omega) d\omega} 
\end{equation}
(in practice, the energy of the modes satisfying the dispersion
relation is computed with a finite width between $\omega_R\pm 3$). For
wavenumbers up to $k_z \approx 30$, modes compatible with waves
concentrate a large fraction of the energy. However, for $k_z > 30$
the fraction of the energy in the waves drops quickly to less than 
$\approx 10\%$. Interestingly, the wavenumber $k_\Omega$ for which
waves become negligible and isotropy is recovered is much larger in
this simulation, $k_\Omega \approx 460$ (see \cite{Clark14a}). In
fact, it can be shown that waves become subdominant in the
spa\-tio-tem\-po\-ral spectrum as soon as the sweeping time becomes of 
the same order as the wave period \cite{Clark14a}. Thus, at moderate
Rossby numbers, eddies (although strongly anisotropic) are relevant for
the energetics of inertial-range rotating turbulence and cannot be
neglected.

To further quantify the contribution of vortical modes and
    of eddies to the energy spectrum, we can estimate a frequency
    spectrum of kinetic energy from the slices of the $E({\bf
      k},\omega)$ spectrum. As an example, from the three-dimensional
    spectrum $E(k_x=0,k_y,k_z,\omega)$, we can estimate a ``total''
    frequency spectrum as
    \begin{align}
        E_{TOT}(\omega) = \sum_{k_y,k_z} E(k_x=0,k_y,k_z,\omega),
        \label{eyz}
    \end{align}
    a ``vortical'' spectrum that has only contributions from the slow
    or vortical modes (i.e., the modes with $k_z=0$, 
    such that the wave frequency is $\omega_R =0$),
    \begin{align}
        E_v(\omega) = \sum_{k_y} E(k_x=0,k_y,k_z=0,\omega),
        \label{ey}
    \end{align}
    and finally, the frequency spectrum of all the modes which are
    often associated with fast or wave modes in wave turbulence
    theories \cite{Smith02,Bourouiba08,Sen12}
    \begin{align}
        E_w(\omega) &= \sum_{k_y,k_z>0} E(k_x=0,k_y,k_z,\omega) 
            \nonumber \\
        &= E_{TOT}(\omega) - E_v(\omega) .
        \label{ez}
    \end{align}
    Note that $E_w(\omega)$, although often associated with the
    spectrum of the waves, contains contributions from modes in
    Fig.~\ref{rotekw} which do not satisfy the dispersion relation
    $\omega_R({\bf k})$.

The resulting spectra are shown in Fig.~\ref{rotew}. The spectrum
    of vortical modes $E_v$ gives the largest contribution to
    $E_{TOT}$ for frequencies $\omega\lesssim4$, where the strong
    two-dimensional modes carry most of the energy. For 
    $4<\omega\le 2\Omega$, $E_w$ becomes dominant and wave modes give
    the largest contribution to the energy. However, for 
    $\omega > 2\Omega$ there are no inertial waves (as the frequency
    of inertial waves has an upper bound of $2\Omega$), and all modes
    that contribute to the spectra must be associated with vortical
    motions (even for $E_w$). All the spectra in this range then show
    the same behavior. The coexistence of multiple timescales in
    this system does not allow a simple transformation of the
    wavenumber spectrum into the frequency spectrum, as was done in
    Sec.~\ref{sec:HIT} for isotropic and homogeneous turbulence.

\begin{figure}
    \centering
    \includegraphics[width=8.5cm]{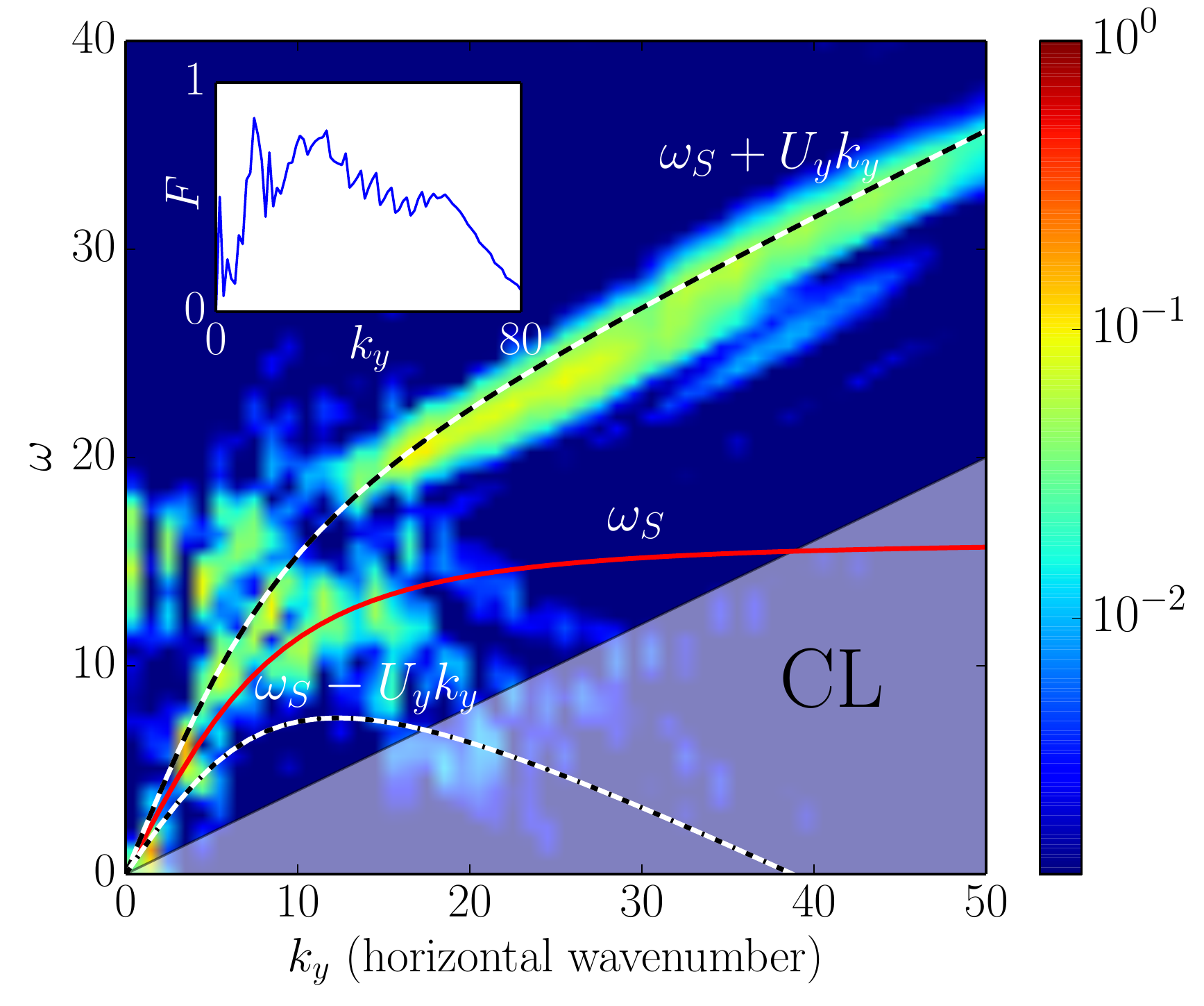}
    \caption{Spatio-temporal spectrum of the potential energy, $E_\theta
        (k_x=0,k_y,k_z=10)$ in the simulation of stratified turbulence.
        The dispersion relation of internal gravity waves, 
        $\omega_S({\bf k})$, is indicated by the solid line. Also shown
        are two Doppler-shifted dispersion relations, corresponding to
        the horizontal r.m.s.~velocities $U_y = \pm 0.4$, and
        indicated by the dashed and dash-dotted lines. Energy is not
        concentrated around the linear dispersion relation, instead it
        is spread in the fan between these two Doppler-shifted
        branches. Moreover, the distribution of energy is not uniform
        as waves with $\omega < U_y k_y$ are absorbed by the mean flow
        in Critical Layers (indicated by the shaded area labeled
        ``CL''). Note the almost complete lack of energy in this
        region. {\it Inset}: fraction of the energy contained within
        the two Doppler shifted branches.}
    \label{strekw}
\end{figure}

\begin{figure}
    \centering
    \includegraphics[width=8cm]{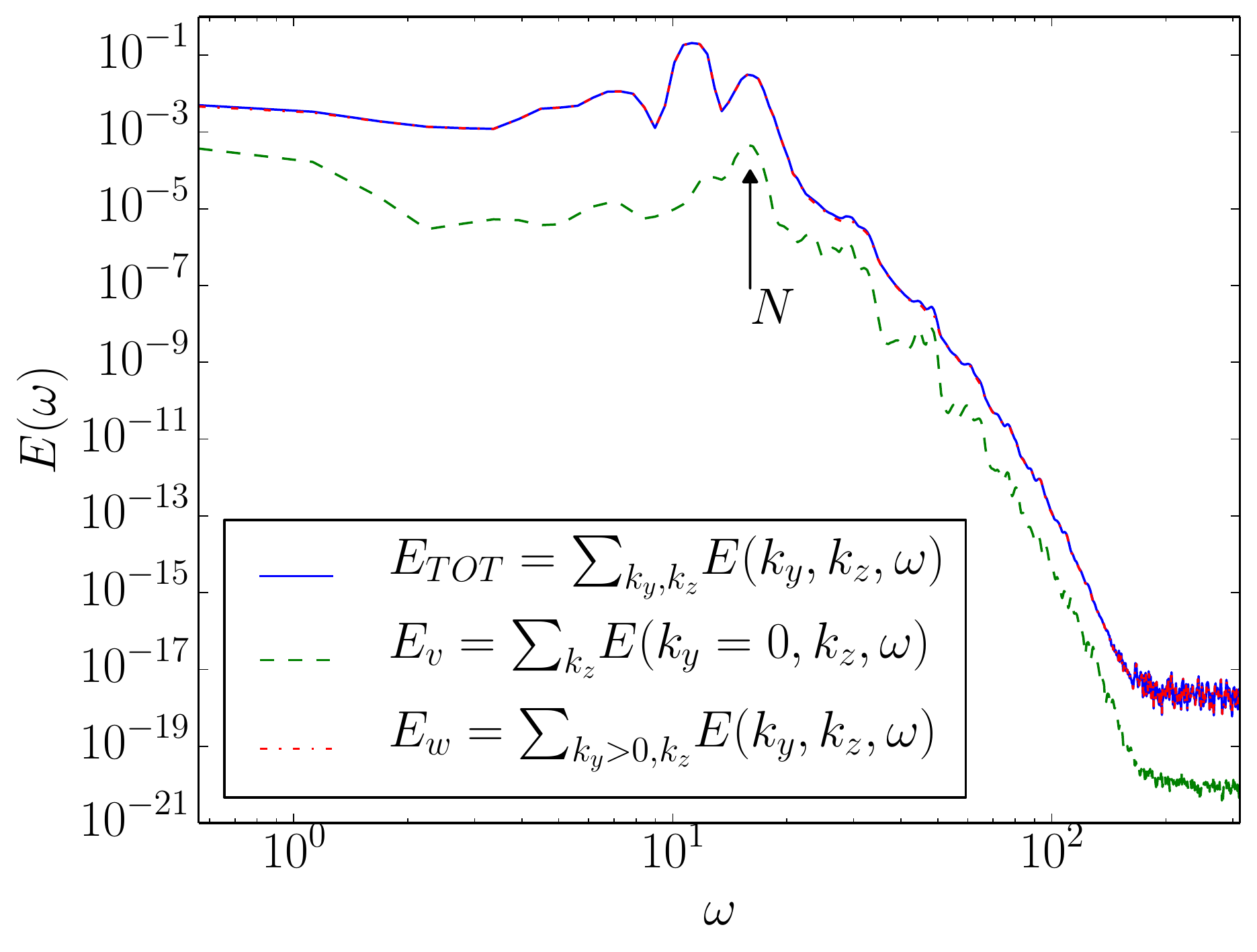}
    \caption{Frequency spectra of potential energy in the simulation
      of stratified turbulence. In this case $E_v$ (dash-dotte red line)
      has the contribution of vortical modes, $E_w$ (dashed
      green line) has the contributions of wave modes, and $E_{TOT}$
      (solid blue line) has the sum of the spatio-temporal spectrum
      over all wavenumbers. Note $E_v$ and $E_{TOT}$ coincide at
      almost all wavenumbers, indicating a dominance of wave
      modes. A peak at frequency $\omega = N$ is indicated by the
      arrow, although in this case frequencies larger than $N$ can
      still be associated with wave motions per virtue of the Doppler
      shift observed in Fig.~\ref{strekw}.}
    \label{staew}
\end{figure}

% Stratified turbulence
\subsection{Stratified turbulence} 

Stratified turbulence shares similarities with the rotating case,
as the buoyancy force gives rise to internal gravity waves, which in
turn create anisotropy as resonant non-linear interactions transfer
energy preferentially to modes with $k_\perp \approx 0$. As a result,
while rotating turbulence develops columnar structures in the vertical 
direction with the fastest waves propagating in that direction,
stratified turbulence generates pancake-like horizontal structures
with the fastest waves propagating horizontally. A related and
well known feature of stratified turbulence is the generation of
large-scale Vertically Sheared Horizontal Winds (VSHW) \cite{Smith02}.  
However, unlike the rotating case where inverse cascades have been
observed for moderate Rossby numbers \cite{Sen12,Alexakis15},
generation of VSHW must have a different origin as inverse cascades
are not possible in stratified flows without the presence of rotation
\cite{Marino13,Herbert14}. In the presence of these winds, waves can
suffer Doppler shift \cite{Hines91}. Also, when in a horizontal layer
the phase velocity of a travelling wave matches the velocity of the
horizontal wind in that layer, the wave is destroyed and its energy
and momentum can be transferred to the mean flow. This phenomenon,
known as critical layer absorption \cite{Hines91,Winters94}, can be
responsible for the generation of the VSHW in stratified
turbulence as shown in  \cite{Clark15}.

\begin{figure}
    \centering
    \includegraphics[width=8.5cm]{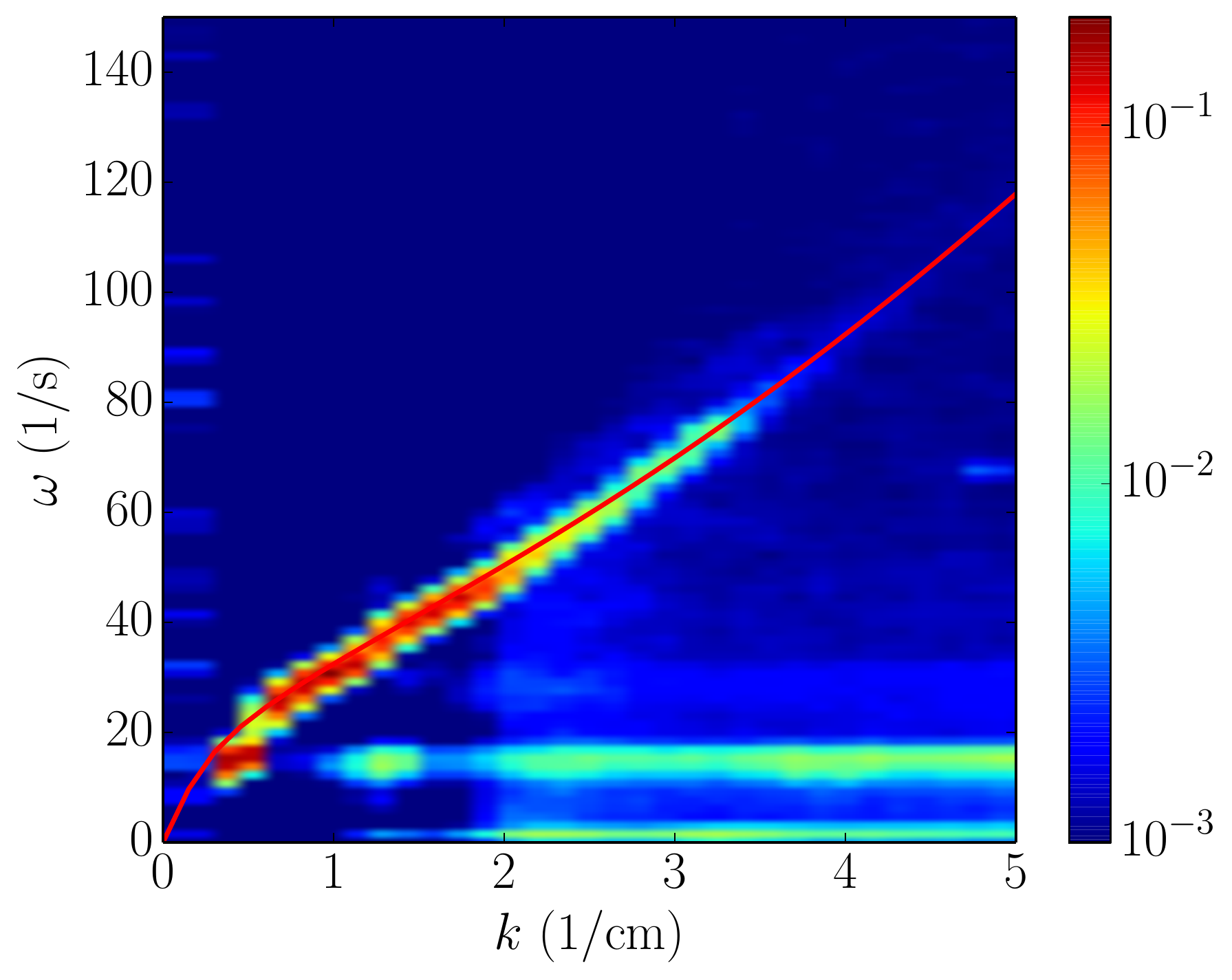}
    \includegraphics[width=8.5cm]{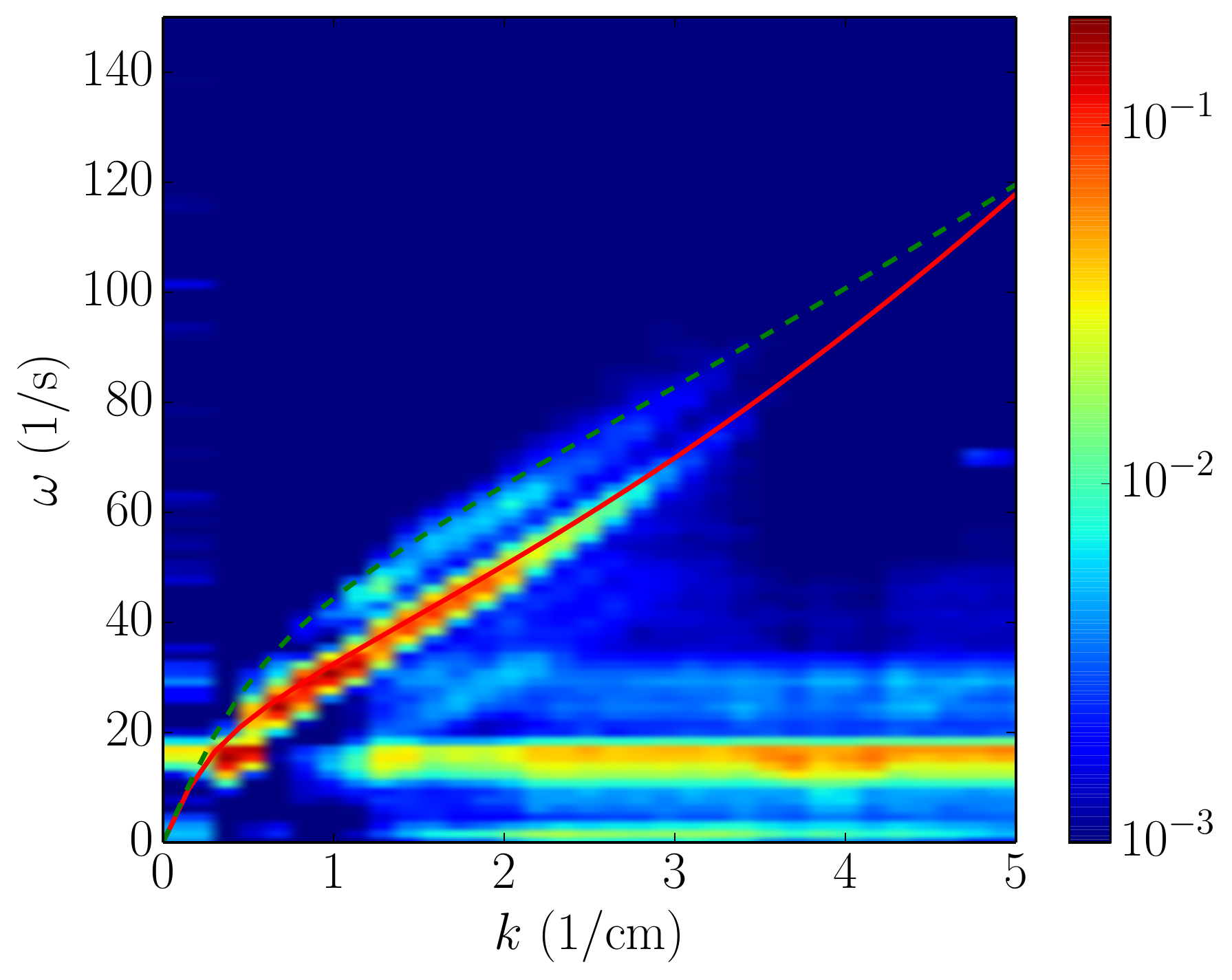}
    \includegraphics[width=8.5cm]{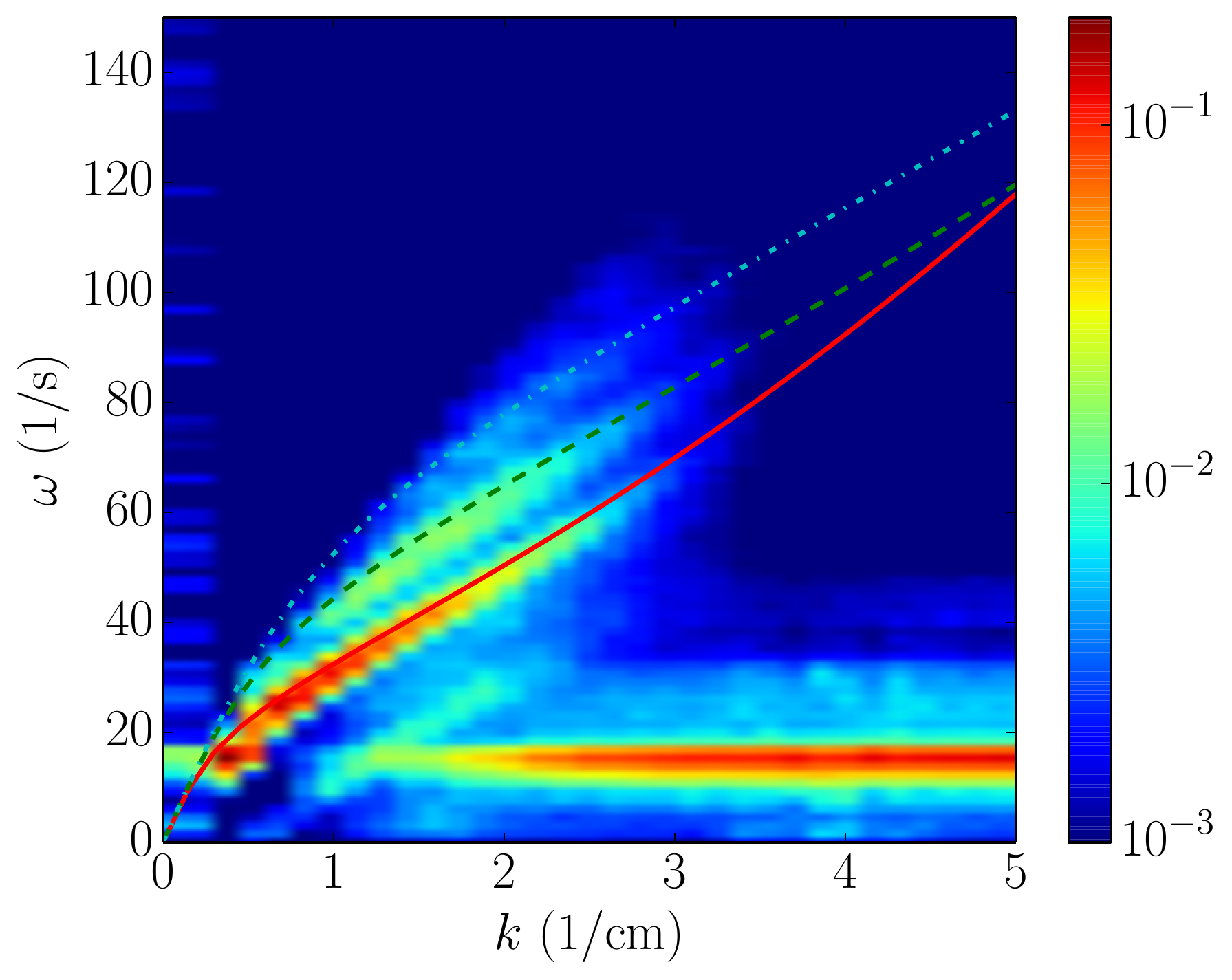}
    \caption{Spatio-temporal power spectrum of the surface deformation
      in CGS units for three surface wave experiments. From top to
      bottom, the amplitude of the forcing is respectively $1$ cm,
      $2$ cm, and $3$ cm. The solid line indicates the linear
      dispersion relation of gravity-capillary waves $\omega_W(k)$, 
      the dashed line indicates the dispersion relation of 2nd-order
      bound waves, and the dash-dotted line indicates the dispersion
      relation of 3rd-order bound waves.}
    \label{surfekw}
\end{figure}

In Fig.~\ref{strekw} we show the spatio-temporal spectrum of the
potential energy, $E_\theta (k_x=0,k_y,k_z=10)$, for the simulation of
stratified turbulence. As in the previous section, we use a flat top
filter in time to compute the spectrum and reduce spectral leakage. 
The dispersion relation of internal gravity waves, $\omega_S(k)$, and
two Doppler shifted branches, $\omega_S(k) + U_y k_y$ and
$\omega_S(k) - U_y k_y$ are also shown as a reference (where $U_y=0.4$
is the horizontal r.m.s.~velocity measured from the flow). Unlike the
rotating case, energy is not concentrated in a narrow region around
the linear dispersion relation. Instead, energy is spread within the
two Doppler-shifted branches. This is the result of the presence of
VSHW in the flow. In each horizontal layer the horizontal velocity
takes values between $\approx \pm U_y$, and the frequency of the waves
is shifted accordingly. Moreover, the distribution of energy in the
fan defined by the two Doppler shifted dispersion relations is not
uniform, as the region with $\omega < U_y k_y$ has almost negligible
energy. This is the result of critical layer absorption: for $\omega <
U_y k_y$, there are layers such that the phase velocity can match the
horizontal wind, and waves are then absorbed.

Although there is more complexity and a larger variety of phenomena
involving waves in this flow than in the rotating case, modes that can
be associated with waves concentrate more energy in a wider range of
scales than in the rotating flow discussed in the previous section. To
illustrate this, Fig.~\ref{strekw} also shows the ratio of the
potential energy in the fan between the two dispersion relations 
$\omega_S (k) \pm U_y k_y$, to the total potential energy at the same
wavenumber. Interestingly, and except at the largest scales (smallest
wavenumbers), around $70\%$ of the energy corresponds to Doppler
shifted waves.

\begin{figure}
    \centering
    \includegraphics[width=8cm]{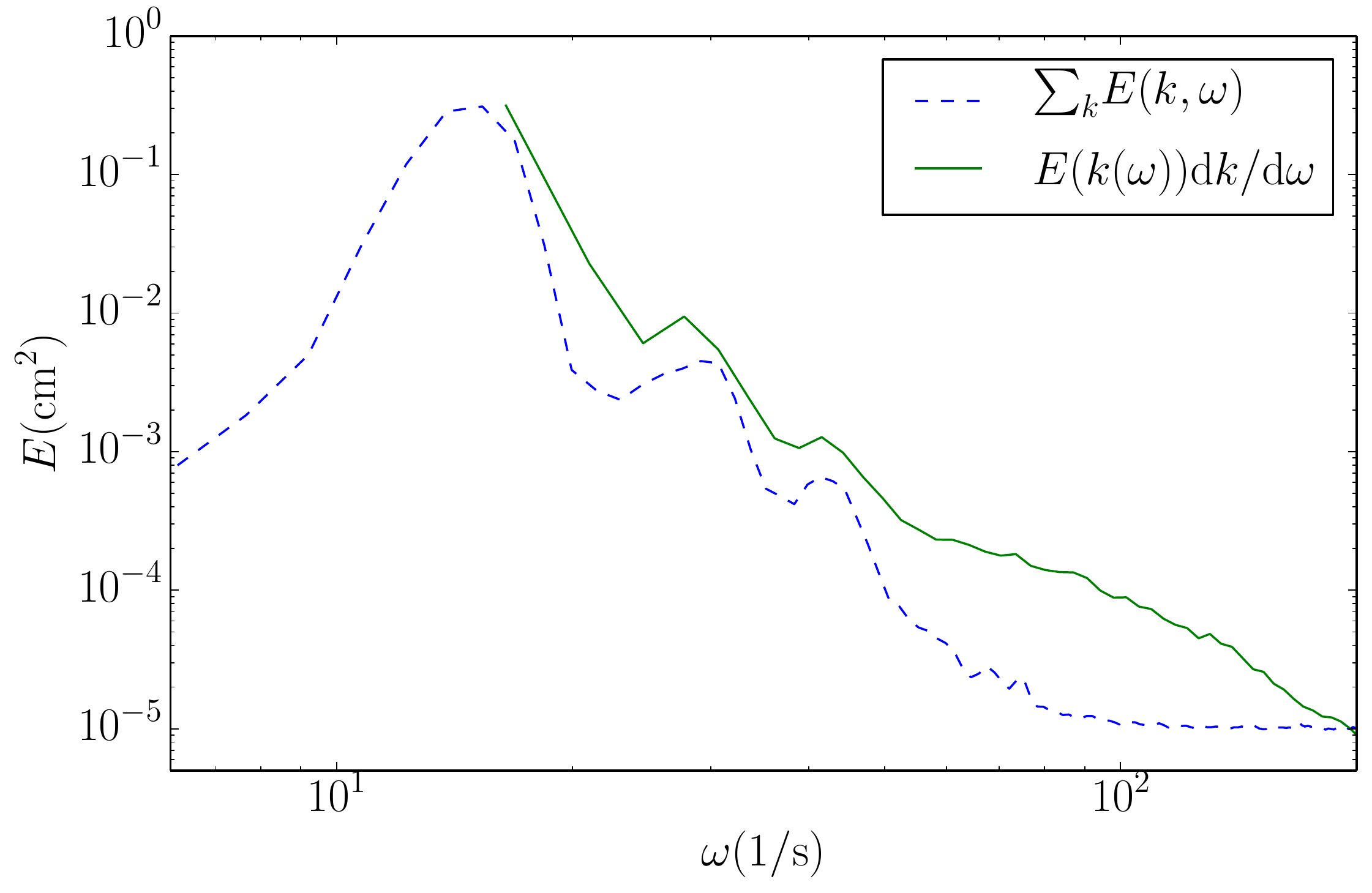}
    \caption{Frequency spectrum $E(\omega)$ from the water waves
        experiment, calculated directly as $\sum_k E(k,\omega)$)
        (dashed blue line), and by changing variables in the
        wavenumber spectrum using the dispersion relation
        $\omega=\omega_W(k)$ (solid green line). As waves provide the
        only relevant timescale in this system, the main features from
        one spectrum can be recovered from the other.}
    \label{labew}
\end{figure}

Following the same procedure as in the case of rotating
  turbulence, we can calculate frequency spectra from the
  spatio-temporal spectrum. Note however that vortical (or slow) modes
  in the stratified case are associated with modes with $k_\perp=0$,
  while wave (or fast) modes now are associated with modes with
  $k_\perp>0$. The resulting spectra are shown in
  Fig.~\ref{staew}. Unlike the rotating case shown in
  Fig.~\ref{rotew}, now most of the energy seems to be in the
  $E_w(\omega)$ spectrum. Moreover, frequencies $\omega > 2N$ now can
  still be associated with wave motions per virtue of the Doppler
  shift observed in Fig.~~\ref{strekw}.

% Surface waves
\subsection{Water waves} 

Water waves are the archetypical wave turbulent system, and have been
studied by such theories ever since their inception (see an
introduction in \cite{Nazarenko}). As a result, determining whether 
the energy of the system is indeed concentrated in wave modes, and
what is the precise spectrum of the waves, were always major
concerns for experimentalists (see, e.g., \cite{Maurel09,Cobelli09}
and references therein). New experimental techniques, such as those
described in the introduction, allowed detailed studies of free-surface
deformation with spatial and temporal resolution, thus allowing direct
computation of the wave energy spectrum without {\it a priori}
knowledge of the dispersion relation of the system.

In Fig.~\ref{surfekw} we show the spatio-temporal power spectrum of
the fluctuations of the surface height deformation $h(x,y,t)$, as
obtained from the experiment using the EMDP technique. This spectrum is
proportional to the spectrum of the potential energy, and under
certain conditions, proportional to the spectrum of the so-called wave
action in wave turbulence theories \cite{Nazarenko}. Three cases are
shown under the three different forcings described above, i.e., for
displacements of the wave makers respectively of $A=1$ cm, $2$ cm, and
$3$ cm.

In all cases shown in Fig.~\ref{surfekw} there is a concentration of
energy around modes that satisfy the dispersion relation of
gravity-capillary surface waves, $\omega_W(k)$, indicated in the
figure by the solid line. There is also a trace of the forcing for
$\omega = 2\pi \times 2.7 \, \textrm{s}^{-1} \approx 17 \,
\textrm{s}^{-1}$, with increasing power as the amplitude of the
forcing is increased. Also, as the forcing increases, new modes that
do not satisfy any of these relations are excited.

Indeed, for larger forcing amplitudes two features can be
identified. On the one hand, the width (in terms of the wavenumber) of
the dispersion relation $\omega_W(k)$ decreases: while for forcing
amplitude of $1$ cm gravity-capillary waves can be identified up to $k
\approx 4$ cm$^{-1}$, for forcing amplitude of $3$ cm the dispersion
relation extends up to $k \approx 3$ cm$^{-1}$. On the other hand, new
wave branches are excited. These correspond to bound waves
\cite{Longuet63,Herbert10}. Bound waves are small amplitude waves that 
travel in front of (i.e., they have the same phase velocity as) a
larger amplitude parent wave. They can be generated by the breaking of
the longer waves, by nonlinear distortion of the parent wave, or be
parasitic capillaries travelling on the front of the parent wave. As
they travel on the front of the long wave, they are phase-locked, and
as they have the same phase velocity, their dispersion relation is
given by \cite{Longuet63,Herbert10}
\begin{align*}
    \Omega_N (N k) = N \omega_W (k) ,
\end{align*}
where $N$ is the order of the bounded waves. The dotted and
dash-dotted lines in Fig.~\ref{surfekw} indicate the dispersion
relation of bound waves for $N=2$ and $3$. The shortening of the width
of the dispersion relation $\omega_W(k)$, together with the excitation
of higher order bound waves as the forcing amplitude is increased,
indicates that with larger amplitudes the energy transfer mechanism
shifts from a cascade to gravity-capillary waves with smaller
wavenumbers, to a transfer of energy towards higher-order bound waves
with smaller frequencies.

For one of the experiments (the case with displacement of the
  wavemakers of 1 cm), we show in Fig.~\ref{labew} the
  frequency spectrum $E(\omega)$ calculated as $\sum_k E(k,\omega)$,
  and the spectrum computed from the spatial spectrum $E(k)$ using the
  dispersion relation to change variables to obtain 
  $E(k(\omega))dk/d\omega$. Although the agreement is not as good as
  in the case of isotropic and homogeneous turbulence
  (Fig.~\ref{hitew}), the main features of one spectrum can be
  recovered from the other. This suggests that the dominant time scale
  in the system is given by the period of the waves, unlike the
  previous cases where multiple time scales could be identified at a
  given wavenumber.

% Conclusions 
\section{Conclusions} 

We presented four examples of turbulent flows where the
spatio-temporal spectrum can be used to identify key flow features and
reveal aspects of their dynamics. The examples considered are
numerical simulations of isotropic and homogeneous turbulence, of
rotating turbulence, and of stratified turbulence, and laboratory
experiments of gravity-capillary water wave turbulence. Two of these
cases were reported extensively in \cite{Clark14a,Clark15}, while the
results for the cases of isotropic and homogeneous turbulence and of
water wave turbulence are new.

In all these cases, the spatio-temporal spectrum allows quantification
of the energy in wave motions, in bound waves, and in eddies, as well
as identification of physical effects such as sweeping, Doppler
shift, and critical layer absorption. For isotropic and homogeneous
turbulence, the dominant Eulerian timescale is sweeping, as
expected from previous studies considering decorrelation times
\cite{Chen89,Nelkin90,Sanada92}. For rotating turbulence, the dominant
timescale at intermediate scales is the period of inertial waves,
although the energy in inertial waves drops to less than 
$\approx 10\%$ for scales which are still much larger than the Zeman
scale for which eddies are expected to be as important as the
waves. The case of stratified turbulence is much more complex, with
the spatio-temporal spectrum showing Doppler shift of internal gravity
waves by the horizontal winds, and absorption of waves in critical
layers where the wave speed matches the horizontal wind
speed. Finally, in the laboratory experiments of water wave
turbulence, the spectrum shows evidence of the presence of bound
waves, as reported before in \cite{Longuet63,Herbert10}. A study
varying the forcing indicates that as the forcing amplitude is
increased, the energy transfer mechanism shifts from a cascade to
gravity-capillary waves with smaller wavenumbers, to a transfer of
energy towards higher-order bound waves with smaller frequencies.

\bibliographystyle{unsrt}
\bibliography{ms}

\begin{thebibliography}{10}

\bibitem{Biferale04}
A.~Celani B. J. Devenish A.~Lanotte L.~Biferale, G.~Boffetta and F.~Toschi.
\newblock Multifractal statistics of lagrangian velocity and acceleration in
  turbulence.
\newblock {\em Phys.\ Rev.\ Lett.}, 93:064502, Aug 2004.

\bibitem{Hussain86}
A.~K. M.~Fazle Hussain.
\newblock Coherent structures and turbulence.
\newblock {\em J.\ Fluid Mech.}, 173:303--356, December 1986.

\bibitem{Rogers87}
M.~M. Rogers and P.~Moin.
\newblock The structure of the vorticity field in homogeneous turbulent flows.
\newblock {\em J.\ Fluid Mech.}, 176:33--66, 1987.

\bibitem{Berkooz93}
P.~Holmes G.~Berkooz and J.~L. Lumley.
\newblock The {Proper} {Orthogonal} {Decomposition} in the {Analysis} of
  {Turbulent} {Flows}.
\newblock {\em Annual Review of Fluid Mechanics}, 25(1):539--575, 1993.

\bibitem{Smith05}
J.~Moehlis T.~R.~Smith and P.~Holmes.
\newblock Low-dimensional modelling of turbulence using the proper orthogonal
  decomposition: a tutorial.
\newblock {\em Nonlinear Dynamics}, 41(1-3):275--307, 2005.

\bibitem{Woods80}
J.~D. Woods.
\newblock Do waves limit turbulent diffusion in the ocean?
\newblock {\em Nature}, 288:219--224, November 1980.

\bibitem{Cambon89}
C.~Cambon and L.~Jacquin.
\newblock Spectral approach to non-isotropic turbulence subjected to rotation.
\newblock {\em J.\ Fluid Mech.}, 202:295--317, 1989.

\bibitem{Waleffe93}
F.~Waleffe.
\newblock Inertial transfers in the helical decomposition.
\newblock {\em Phys.\ Fluids A}, 5(3):677, 1993.

\bibitem{Cambon97}
C.~Cambon, N.~N. Mansour, and F.~S. Godeferd.
\newblock Energy transfer in rotating turbulence.
\newblock {\em J.\ Fluid Mech.}, 337:303--332, 1997.

\bibitem{Nazarenko}
Sergey Nazarenko.
\newblock {\em Wave Turbulence}.
\newblock Springer, 2011 edition, February 2011.

\bibitem{Stewart69}
R.~W. Stewart.
\newblock Turbulence and {Waves} in a {Stratified} {Atmosphere}.
\newblock {\em Radio Science}, 4:1269--1278, December 1969.

\bibitem{Smith02}
L.~M. Smith and F.~Waleffe.
\newblock Generation of slow large scales in forced rotating stratified
  turbulence.
\newblock {\em J.\ Fluid Mech.}, 451:145--168, January 2002.

\bibitem{Bourouiba08}
L.~Bourouiba.
\newblock Model of a truncated fast rotating flow at infinite reynolds number.
\newblock {\em Phys.\ Fluids}, 20(7):--, 2008.

\bibitem{Sen12}
D.~Rosenberg A.~Sen, P. D.~Mininni and A.~Pouquet.
\newblock Anisotropy and nonuniversality in scaling laws of the large-scale
  energy spectrum in rotating turbulence.
\newblock {\em Phys.\ Rev.\ E}, 86:036319, Sep 2012.

\bibitem{Adrian91}
R.~J. Adrian.
\newblock Particle-imaging techniques for experimental fluid mechanics.
\newblock {\em Annual review of fluid mechanics}, 23(1):261--304, 1991.

\bibitem{Maurel09}
V.~Pagneux A.~Maurel, P.~Cobelli and P.~Petitjeans.
\newblock Experimental and theoretical inspection of the phase-to-height
  relation in {Fourier} transform profilometry.
\newblock {\em Applied Optics}, 48(2):380--392, January 2009.

\bibitem{Cobelli09}
A.~Maurel V.~Pagneux P.~Cobelli, P.~Petitjeans and N.~Mordant.
\newblock Space-{Time} {Resolved} {Wave} {Turbulence} in a {Vibrating} {Plate}.
\newblock {\em Phys.\ Rev.\ Lett.}, 103(20):204301, November 2009.

\bibitem{Lagubeau15}
G.~Lagubeau, P.~Cobelli, T.~Bobinski, A.~Maurel, V.~Pagneux, and P.~Petitjeans.
\newblock Empirical mode decomposition profilometry: small-scale capabilities
  and comparison to fourier transform profilometry.
\newblock {\em Appl. Opt.}, 54(32):9409--9414, Nov 2015.

\bibitem{Liu12}
P.~Carns C. Carothers R. Ross G. Grider A.~Crume N.~Liu, J.~Cope and
  C.~Maltzahn.
\newblock On the role of burst buffers in leadership-class storage systems.
\newblock In {\em Mass Storage Systems and Technologies (MSST), 2012 IEEE 28th
  Symposium on}, pages 1--11. IEEE, 2012.

\bibitem{Campagne15}
F.~Moisy A.~Campagne, B.~Gallet and P.-P. Cortet.
\newblock Disentangling inertial waves from eddy turbulence in a forced
  rotating turbulence experiment.
\newblock {\em Phys.\ Rev.\ E}, April 2015.

\bibitem{Bewley07}
Gregory~P. Bewley, Daniel~P. Lathrop, Leo R.~M. Maas, and K.~R. Sreenivasan.
\newblock Inertial waves in rotating grid turbulence.
\newblock {\em Phys.\ Fluids}, 19(7):071701, July 2007.

\bibitem{Rieutord12}
D.~S.~Zimmerman M.~Rieutord, S. A.~Triana and D.~P. Lathrop.
\newblock Excitation of inertial modes in an experimental spherical {Couette}
  flow.
\newblock {\em Phys.\ Rev.\ E}, 86(2):026304, August 2012.

\bibitem{Lamriben11}
Cyril Lamriben, Pierre-Philippe Cortet, Fr\'ed\'eric Moisy, and Leo R.~M. Maas.
\newblock Excitation of inertial modes in a closed grid turbulence experiment
  under rotation.
\newblock {\em Phys.\ Fluids}, 23(1):015102, January 2011.

\bibitem{Servidio11}
P.~Dmitruk S.~Servidio, V.~Carbone and W.~H. Matthaeus.
\newblock Time decorrelation in isotropic magnetohydrodynamic turbulence.
\newblock {\em Europhys.\ Lett.}, 96(5):55003, December 2011.

\bibitem{Favier10}
F.~S.~Godeferd B.~Favier and C.~Cambon.
\newblock On space and time correlations of isotropic and rotating turbulence.
\newblock {\em Phys.\ Fluids}, 22(1):015101, 2010.

\bibitem{Clark14a}
P.~D. Mininni P.~Dmitruk P.~Clark~di Leoni, P. J.~Cobelli and W.~H. Matthaeus.
\newblock Quantification of the strength of inertial waves in a rotating
  turbulent flow.
\newblock {\em Phys.\ Fluids}, 26(3):035106, March 2014.

\bibitem{Dmitruk09}
P.~Dmitruk and W.~H. Matthaeus.
\newblock Waves and turbulence in magnetohydrodynamic direct numerical
  simulations.
\newblock {\em Phys.\ Plasmas}, 16(6):062304, June 2009.

\bibitem{Lindborg07}
E.~Lindborg and G.~Brethouwer.
\newblock Stratified turbulence forced in rotational and divergent modes.
\newblock {\em J.\ Fluid Mech.}, 586:83--108, September 2007.

\bibitem{Dasaro00}
E.~A. D’Asaro and R.-C. Lien.
\newblock Lagrangian {Measurements} of {Waves} and {Turbulence} in {Stratified}
  {Flows}.
\newblock {\em Journal of Physical Oceanography}, 30(3):641--655, March 2000.

\bibitem{Cobelli11}
P.~Cobelli, A.~Przadka, P.~Petitjeans, G.~Lagubeau, V.~Pagneux, and A.~Maurel.
\newblock Different {Regimes} for {Water} {Wave} {Turbulence}.
\newblock {\em Phys.\ Rev.\ Lett.}, 107(21):214503, November 2011.

\bibitem{Aubourg15}
Quentin Aubourg and Nicolas Mordant.
\newblock Nonlocal {Resonances} in {Weak} {Turbulence} of {Gravity}-{Capillary}
  {Waves}.
\newblock {\em Phys.\ Rev.\ Lett.}, 114:144501, April 2015.

\bibitem{Clark14b}
P.~D.~Mininni P.~Clark~di Leoni, P. J.~Cobelli.
\newblock Wave turbulence in shallow water models.
\newblock {\em Phys.\ Rev.\ E}, 89(6):063025, June 2014.

\bibitem{Yokoyama14}
Naoto Yokoyama and Masanori Takaoka.
\newblock Identification of a separation wave number between weak and strong
  turbulence spectra for a vibrating plate.
\newblock {\em Phys.\ Rev.\ E}, 89(1):012909, January 2014.

\bibitem{Yarom14}
E.~Yarom and E.~Sharon.
\newblock Experimental observation of steady inertial wave turbulence in deep
  rotating flows.
\newblock {\em Nature Physics}, 10(7):510--514, June 2014.

\bibitem{Meyrand15}
R.~Meyrand, K.~H. Kiyani, and S.~Galtier.
\newblock Weak magnetohydrodynamic turbulence and intermittency.
\newblock {\em J.\ Fluid Mech.}, 770, May 2015.

\bibitem{Clark15}
P.~Clark di~Leoni and P.~D. Mininni.
\newblock Absorption of waves by large-scale winds in stratified turbulence.
\newblock {\em Phys.\ Rev.\ E}, 91(3):033015, March 2015.

\bibitem{Nazarenko06a}
Sergey Nazarenko and Miguel Onorato.
\newblock Wave turbulence and vortices in {Bose}–{Einstein} condensation.
\newblock {\em Physica D: Nonlinear Phenomena}, 219(1):1--12, July 2006.

\bibitem{Nazarenko06b}
Sergey Nazarenko and Miguel Onorato.
\newblock Freely decaying {Turbulence} and {Bose}–{Einstein} {Condensation}
  in {Gross}–{Pitaevski} {Model}.
\newblock {\em Journal of Low Temperature Physics}, 146(1-2):31--46, January
  2007.

\bibitem{Clark15b}
Patricio Clark~di Leoni, Pablo~D. Mininni, and Marc~E. Brachet.
\newblock Direct evidence of {Kelvin} waves in numerical simulations of quantum
  turbulence.
\newblock {\em arXiv:1509.05316 [physics]}, September 2015.
\newblock arXiv: 1509.05316.

\bibitem{Sahraoui10}
M.~L.~Goldstein F.~Sahraoui, G.~Belmont and L.~Rezeau.
\newblock {Limitations of multispacecraft data techniques in measuring wave
  number spectra of space plasma turbulence}.
\newblock {\em Journal of Geophysical Research}, 115(A4):A04206--10, April
  2010.

\bibitem{Sahraoui11}
G.~Belmont A. Roux L. Rezeau P. Canu P. Robert N. Cornilleau-Wehrlin O. Le
  Contel T. D. De Wit J.-L.~Pin{\c c}on F.~Sahraoui, M. L.~Goldstein and
  K.~Kiyani.
\newblock {Multi-spacecraft investigation of space turbulence Lessons from
  Cluster and input to the Cross-Scale mission}.
\newblock {\em Planetary and Space Science}, 59(7):585--591, May 2011.

\bibitem{Escoubet}
C.P. Escoubet, R.~Schmidt, and C.~Russell.
\newblock {\em The Cluster and Phoenix Missions}.
\newblock Space science reviews. Springer Netherlands, 1997.

\bibitem{Morton15}
S.~Tomczyk R.~J.~Morton and R.~Pinto.
\newblock Investigating {Alfvénic} wave propagation in coronal open-field
  regions.
\newblock {\em Nature Comm.}, 6, 2015.

\bibitem{Gomez05a}
P.~D.~Mininni D.~G\'omez and P.~Dmitruk.
\newblock {MHD} simulations and astrophysical applications.
\newblock {\em Advances in Space Research}, 35(5):899--907, 2005.

\bibitem{Gomez05b}
P.~D.~Mininni D.~G\'omez and P.~Dmitruk.
\newblock Parallel simulations in turbulent {MHD}.
\newblock {\em Phys. Scripta}, 2005:123, 2005.

\bibitem{Mininni11}
R.~Reddy P.~Mininni, D.~Rosenberg and A.~Pouquet.
\newblock A hybrid {MPI-OpenMP} scheme for scalable parallel pseudospectral
  computations for fluid turbulence.
\newblock {\em Parallel Computing}, 37(6-7):316--326, 2011.

\bibitem{Przadka2012}
A.~Przadka, B.~Cabane, V.~Pagneux, A.~Maurel, and P.~Petitjeans.
\newblock Fourier transform profilometry for water waves: how to achieve clean
  water attenuation with diffusive reflection at the water surface?
\newblock {\em Exp. Fluids}, 52(2):519--527, 2012.

\bibitem{Davidson}
P.~A. Davidson.
\newblock {\em Turbulence: an introduction for scientists and engineers}.
\newblock Oxford Univ.\ Press, Oxford, 2004.

\bibitem{Tennekes75}
H.~Tennekes.
\newblock Eulerian and {Lagrangian} time microscales in isotropic turbulence.
\newblock {\em J.\ Fluid Mech.}, 67(3):561--567, 1975.

\bibitem{Chen89}
S.~Chen and R.~H. Kraichnan.
\newblock Sweeping decorrelation in isotropic turbulence.
\newblock {\em Phys.\ Fluids A}, 1(12):2019, December 1989.

\bibitem{Nelkin90}
M.~Nelkin and M.~Tabor.
\newblock Time correlations and random sweeping in isotropic turbulence.
\newblock {\em Phys.\ Fluids A}, 2(1):81--83, January 1990.

\bibitem{Sanada92}
T.~Sanada and V.~Shanmugasundaram.
\newblock Random sweeping effect in isotropic numerical turbulence.
\newblock {\em Phys.\ Fluids A}, 4(6):1245--1250, June 1992.

\bibitem{Mininni12}
D.~Rosenberg P.~D.~Mininni and A.~Pouquet.
\newblock Isotropization at small scales of rotating helically driven
  turbulence.
\newblock {\em J.\ Fluid Mech.}, 699:263--279, 2012.

\bibitem{Newell69}
A.~C. Newell.
\newblock Rossby wave packet interactions.
\newblock {\em J.\ Fluid Mech.}, 35(02):255--271, January 1969.

\bibitem{Cambon04}
C.~Cambon, R.~Rubinstein, and F.~S. Godeferd.
\newblock Advances in wave turbulence: rapidly rotating flows.
\newblock {\em New J.\ Phys.}, 6:73, 2004.

\bibitem{Alexakis15}
A.~Alexakis.
\newblock Rotating taylor--green flow.
\newblock {\em J.\ Fluid Mech.}, 769:46--78, 2015.

\bibitem{Marino13}
D.~Rosenberg R.~Marino, P. D.~Mininni and A.~Pouquet.
\newblock Inverse cascades in rotating stratified turbulence: Fast growth of
  large scales.
\newblock {\em Europhys.\ Lett.}, 102(4):44006, 2013.

\bibitem{Herbert14}
A.~Pouquet C.~Herbert and R.~Marino.
\newblock Restricted equilibrium and the energy cascade in rotating and
  stratified flows.
\newblock {\em J.\ Fluid Mech.}, 758:374--406, November 2014.

\bibitem{Hines91}
C.~O. Hines.
\newblock The saturation of gravity waves in the middle atmosphere. part {II:}
  development of doppler-spread theory.
\newblock {\em J.\ Atmos.\ Sci.}, 48(11):1361--1379, June 1991.

\bibitem{Winters94}
K.~B. Winters and E.~A. D’Asaro.
\newblock Three-dimensional wave instability near a critical level.
\newblock {\em J.\ Fluid Mech.}, 272:255--284, August 1994.

\bibitem{Longuet63}
M.~S. Longuet-Higgins.
\newblock The generation of capillary waves by steep gravity waves.
\newblock {\em J.\ Fluid Mech.}, 16(01):138--159, 1963.

\bibitem{Herbert10}
N.~Mordant E.~Herbert and E.~Falcon.
\newblock Observation of the {Nonlinear} {Dispersion} {Relation} and {Spatial}
  {Statistics} of {Wave} {Turbulence} on the {Surface} of a {Fluid}.
\newblock {\em Phys.\ Rev.\ Lett.}, 105(14):144502, September 2010.

\end{thebibliography}

\end{document}